\documentclass[twocolumn,pra,amsmath,amssymb]{revtex4-1}
\usepackage[latin9]{inputenc}
\usepackage{amsmath}
\usepackage{amssymb}
\usepackage{graphicx}
\usepackage[unicode=true,pdfusetitle,
 bookmarks=true,bookmarksnumbered=false,bookmarksopen=false,
 breaklinks=true,pdfborder={0 0 1},backref=false,colorlinks=false]
 {hyperref}
\begin{document}
\title{Emergent Mott insulators and non-Hermitian conservation laws in an
interacting bosonic chain with noninteger filling and nonreciprocal
hopping}
\author{Zuo Wang}
\affiliation{Guangdong Provincial Key Laboratory of Quantum Engineering and Quantum
Materials, School of Physics and Telecommunication Engineering, South
China Normal University, Guangzhou 510006, China}
\author{Li-Jun Lang}
\email{ljlang@scnu.edu.cn}

\affiliation{Guangdong Provincial Key Laboratory of Quantum Engineering and Quantum
Materials, School of Physics and Telecommunication Engineering, South
China Normal University, Guangzhou 510006, China}
\author{Liang He}
\email{liang.he@scnu.edu.cn}

\affiliation{Guangdong Provincial Key Laboratory of Quantum Engineering and Quantum
Materials, School of Physics and Telecommunication Engineering, South
China Normal University, Guangzhou 510006, China}
\begin{abstract}
We investigate the ground state and quantum dynamics of an interacting
bosonic chain with the nonreciprocal hopping. In sharp contrast to
its Hermitian counterpart, the ground state can support Mott insulators
in systems with noninteger filling due to the competition between
nonreciprocal hopping and the on-site interaction. For the quantum
dynamics, conservation laws for non-Hermitian systems manifest a stark
difference from their Hermitian counterpart. In particular, for any
Hermitian operator that commutes with the Hamiltonian operator, its
expectation value is guaranteed to be nonconserved in the non-Hermitian
quantum dynamics. To systematically identify the non-Hermitian conservation
law, we establish a generic approach for constructing the conserved
quantities in non-Hermitian many-body quantum systems with completely
real spectra, and illustrate it concretely by the system under study.
The direct experimental observation of Mott insulators in systems
with noninteger filling and non-Hermitian conservation laws can be
performed by ultracold atoms in optical lattices with the engineered
nonreciprocal hopping.
\end{abstract}
\maketitle

\section{Introduction}

In a very short time, the study of non-Hermitian systems has spread
over almost all branches of physics in recent years, including both
quantum and classical ones \citep{Ashida_Adv_Phy_2020}. Besides the
reason that the gain/loss is naturally inherited in open systems \citep{Breuer_Book_OQS_2002,Daley_Adv_Phy_2014},
the introduction of non-Hermiticity can also give birth to exotic
phenomena that cannot exist in Hermitian counterparts such as the
spontaneous breaking of parity-time symmetry \citep{Bender_PRL_1998}
and exceptional topology \citep{Bergholtz_RMP_2021}. Among them,
one of the most intriguing discussions is about the breakdown of the
bulk-boundary correspondence (BBC) in non-Hermitian topological systems,
which has stimulated many efforts in the construction of topological
theories to rescue it \citep{180720-9,Zhu14,171012-1,180720-8,180720-7}.
Around 2018, many research groups paid special attention to a Su-Schrieffer-Heeger-like
model \citep{Yao_PRL_2018,181016-1,Lieu18,180720-3,180720-6,180720-4,Torres18},
where a so-called \textit{non-Hermitian skin effect} (NHSE) \citep{Yao_PRL_2018}
that assembles bulk states via nonreciprocal hopping to localize at
one end, was found to be responsible for the breakdown of BBC. This
effect is unique to non-Hermitian systems and has been observed in
diverse experimental platforms including photonics \citep{Xue2020,Weidemann2020,Wang_Science_2021},
electrical circuits \citep{Helbig_Nat_Phy_2020,Hofmann_PRR_2020},
mechanical systems \citep{Brandenbourger2019,Ghatak2020}, nitrogen-vacancy
centers \citep{ZhangDuan2021}, and cold atoms \citep{Yan2020}.

At single-particle levels, while NHSEs have been widely studied to
generalize basic notions of the topology in non-Hermitian systems
\citep{Ashida_Adv_Phy_2020}, they also affect prototypical quantum
phenomena such as Anderson localization in non-Hermitian quasicrystals
\citep{190802-1,Longhi2019,Xu2020}. In comparison, at many-body levels,
the discussion of NHSEs on many-body phenomena has just begun to draw
attention \citep{Sato2021,Yoshida2021,Shen_arXiv_2021,Hatsugai2021,Chen_arXiv_2021}.
First attempts have been made at the celebrated one-dimensional Hubbard-like
models to understand how the on-site interaction interplays with the
NHSE \citep{Mu_PRB_2020,Zhang_PRB_2020,Xu_PRB_2020,Lee_PRB_2020,Liu_PRB_2020}.
For fermionic models, the Fermi-Dirac statistics intrinsically prohibits
the nonreciprocal-hopping-induced aggregation of fermions at one end,
resulting in a Fermi-surface-like plateau in real space \citep{Mu_PRB_2020,Lee_PRB_2020};
for bosonic models, the on-site repulsive interaction can still drive
a superfluid--Mott-insulator transition, but the superfluid is skinned
to one end \citep{Zhang_PRB_2020}. In spite of this progress, open
questions are still left, for instance, it is well known that Mott
insulators can only exist at integer fillings for the Hermitian Bose-Hubbard
model \citep{Fisher_PRB_1989,Sachdev_Book_QPT_2011}, therefore a
natural question that arises in this context is whether the interplay
between non-Hermitian nonreciprocal hopping and the interaction could
give rise to a Mott insulator in systems with noninteger filling.
As a matter of fact, even more questions are left open concerning
the quantum many-body dynamics of these interacting non-Hermitian
systems. A particularly fundamental one is the existence and identification
of the conservation laws in non-Hermitian interacting quantum many-body
systems, which not only plays a crucial role in simplifying and solving
physical problems, but also influences strongly the macroscopic universal
dynamical properties of the systems in the vicinities of possible
continuous phase transitions via ``slow-modes'' \citep{Hohenberg_RMP_1977}.

In this paper, we address these questions for an interacting bosonic
chain with nonreciprocal hopping and on-site repulsive interactions.
To this end, we investigate both its ground state and quantum dynamical
properties via exact diagonalization combined with analytic analyses,
and find that the system manifests both distinct static (cf.~Fig.~\ref{fig:GS_properties})
and dynamical (cf.~Figs.~\ref{fig:Dynamics} and \ref{fig:Exp_observability})
properties that are in sharp contrast to those of its Hermitian counterpart.
More specifically, we find the following.

(i) Emergent Mott insulators in systems with noninteger filling due
to the competition between nonreciprocal hopping and the on-site interaction
(cf.~Fig.~\ref{fig:GS_properties}), in sharp contrast to the system's
Hermitian counterpart where Mott insulator can only exist at integer
filling: Due to nonreciprocal hopping, bosons aggregated near one
end of the chain {[}cf.~the red curve of Fig.~\ref{fig:GS_properties}(a){]}
could strongly repel each other in the presence of strong enough on-site
interactions, giving rise to a Mott insulator region with strongly
depressed particle number fluctuations {[}cf.~the black and green
curves of Figs.~\ref{fig:GS_properties}(a) and \ref{fig:GS_properties}(b)
and also Figs.~\ref{fig:GS_properties}(c) and \ref{fig:GS_properties}(d){]}.
Moreover, this result is consistent with that of the noninteracting
Fermi-Hubbard model with nonreciprocal hopping \citep{Mu_PRB_2020,Lee_PRB_2020},
which corresponds to the hard-core limit of the system investigated
here.

(ii) Non-Hermitian conservation laws of the system {[}cf.~Eqs.~(\ref{eq:CQ_deformed_N})--(\ref{eq:CQ_deformed_I})
and Figs.~\ref{fig:Dynamics} and \ref{fig:Exp_observability}{]}
and a systematic approach to identify the non-Hermitian conservation
laws in non-Hermitian many-body quantum systems with completely real
spectra {[}cf.~Eq.~(\ref{eq:non-Hermitian_conservation_laws}){]}:
We find that the conservation laws for the non-Hermitian system generically
manifest a stark difference from their Hermitian counterpart, in particular,
for any Hermitian operator that commutes with the Hamiltonian operator,
its expectation value is guaranteed to be \emph{nonconserved} in the
non-Hermitian quantum dynamics. This clarifies a fundamental concept
that is widely misused in the literature that the total particle number
of this system is conserved because the corresponding operator $\hat{N}$
commutes with the system\textquoteright s Hamiltonian, i.e., $[\hat{N},\hat{H}]=0$
{[}cf.~Figs.~\ref{fig:Dynamics}(a), \ref{fig:Dynamics}(b) and
\ref{fig:Exp_observability} for the time evolution of the expectation
value of the $\hat{N}$, clearly manifesting time dependence{]}. By
employing similarity transformations between the non-Hermitian Hamiltonians
and the auxiliary Hermitian Hamiltonians, we establish a generic approach
for identifying the conservation laws in non-Hermitian many-body quantum
systems with completely real spectra {[}cf.~Eq.~(\ref{eq:non-Hermitian_conservation_laws}){]},
and find a set of conservation laws of the system {[}cf.~Eqs.~(\ref{eq:CQ_deformed_N})--(\ref{eq:CQ_deformed_I})
and Figs.~\ref{fig:Dynamics} and \ref{fig:Exp_observability}{]}.

\section{System and Model}

The system under consideration is an interacting bosonic chain with
nonreciprocal hopping, which is described by the Hamiltonian 
\begin{equation}
\hat{H}=-J\sum_{j=1}^{L-1}(e^{\alpha}\hat{b}_{j}^{\dagger}\hat{b}_{j+1}+e^{-\alpha}\hat{b}_{j+1}^{\dagger}\hat{b}_{j})+\frac{U}{2}\sum_{j}\hat{n}_{j}(\hat{n}_{j}-1),\label{eq:Hamiltonian}
\end{equation}
where $\hat{b}_{j}^{\dagger}$ ($\hat{b}_{j}$) is the creation (annihilation)
operator at site $j$ in the Wannier representation, $\hat{n}_{j}\equiv\hat{b}_{j}^{\dagger}\hat{b}_{j}$
is the particle number operator, and $L$ is the total number of the
lattice sites. The first term in Eq.~(\ref{eq:Hamiltonian}) describes
the nonreciprocal hopping with $J>0$ being the geometric mean and
$\alpha\in\Re$ characterizing the asymmetry of the hopping amplitudes
in opposite directions. The second term describes the on-site interaction
with strength $U$. In general, a non-Hermitian system has a complex
spectrum and thus cannot reach a dynamical equilibrium; here, we focus
on the system with the open boundary condition whose spectrum is completely
real \citep{Zhang_PRB_2020}. Most numerical results to be presented
in the following are obtained by directly diagonalizing the system's
Hamiltonian (\ref{eq:Hamiltonian}) at a finite size $L$ ($L=12$),
and without loss of generality we set $\alpha\ge0$, that is, the
left-biased hopping is considered.

When the hopping is reciprocal ($\alpha=0$), the model Hamiltonian
reduces to the conventional Bose-Hubbard model, where it is well known
that at large enough on-site interaction $U$ the system supports
a Mott insulator only at an integer filling, while a superfluid at
a noninteger filling. Once $\alpha\neq0$, the nonreciprocity of the
hopping is introduced and makes the Hamiltonian non-Hermitian. Previous
investigations in related noninteracting systems have revealed that
nonreciprocal hopping generally can result in NHSEs \citep{Yao_PRL_2018},
where particles tend to accumulate on one end of the chain due to
the unbalanced hopping in opposite directions. In this regard, one
naturally expects the competition between nonreciprocal hopping and
the on-site interaction could give rise to physics that is absent
in conventional Hermitian Bose-Hubbard-type systems and the noninteracting
non-Hermitian bosonic chain. Indeed, as we shall see in the following,
Mott insulators can emerge even in systems with noninteger filling
due to the collaboration between nonreciprocal hopping and the on-site
interaction.

\section{Emergent Mott insulators in systems with noninteger filling}

\begin{figure}
~\includegraphics[width=1.55in]{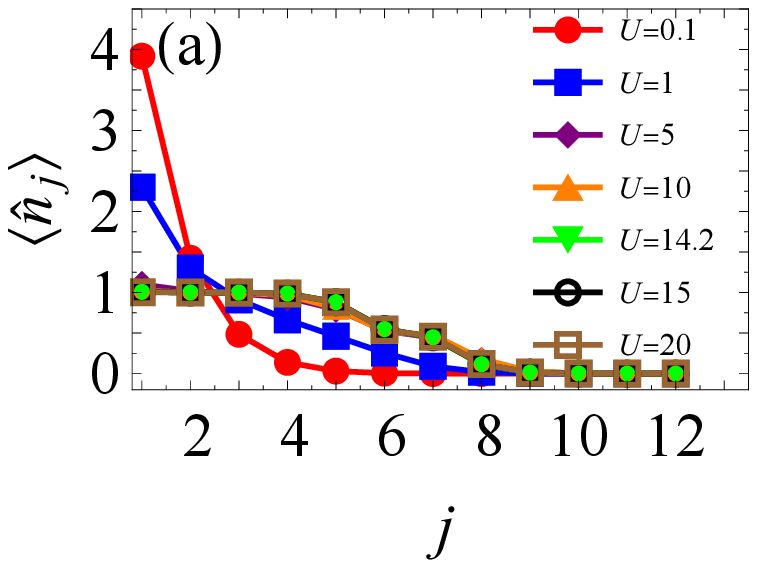}~\includegraphics[width=1.65in]{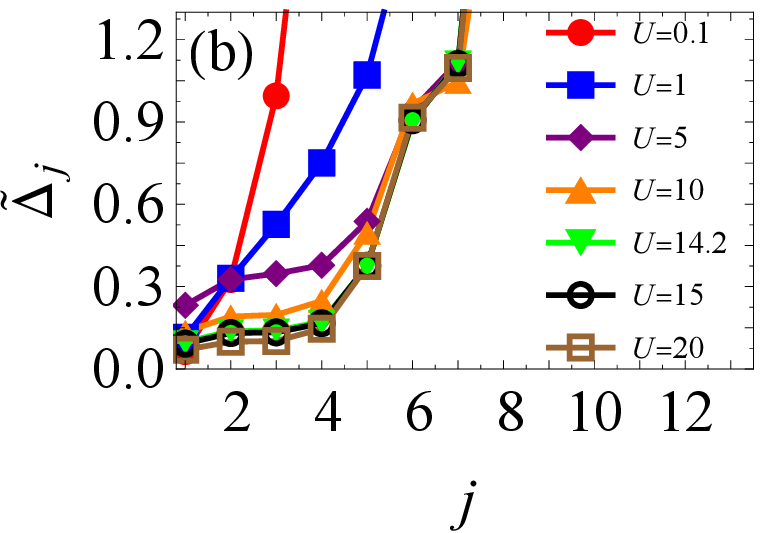}

\includegraphics[width=1.65in]{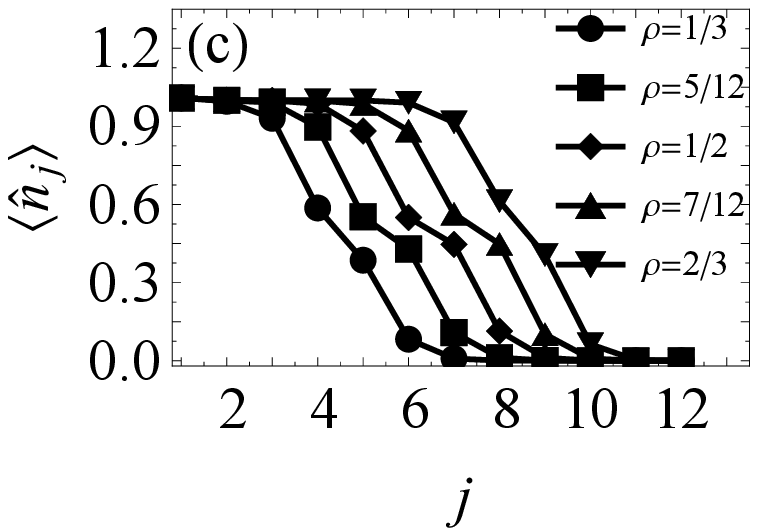}~\includegraphics[width=1.65in]{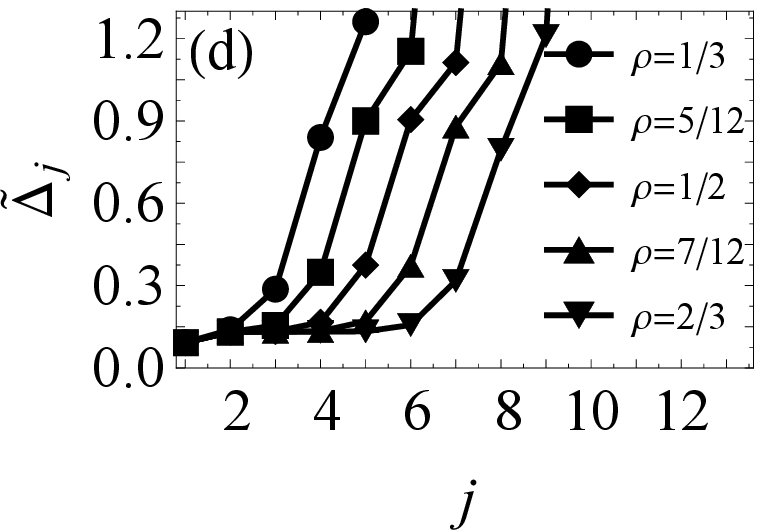}

\caption{Ground state properties of the system. $\alpha=2$, $Je^{\alpha}=1$,
$L=12$ are kept fixed. (a) Real space density distribution $\langle\hat{n}_{j}\rangle$
of a half-filled system at different interaction strengths with $U=0.1,\,1,\,5,\,10,\,14.2,\,15,\,20$.
At a weak interaction strength, the density distribution manifests
a strong NHSE with particles accumulated at the left end (red curve),
while at a strong interaction strength, one can notice a density plateau
emerge (black and brown curves). (b) Corresponding local relative
density fluctuation distributions $\widetilde{\Delta}_{j}$ with $U=0.1,\,1,\,5,\,10,\,14.2,\,15,\,20$.
From the $\widetilde{\Delta}_{j}$ at $U=14.2,\,15,\,20$, one can
notice that in the region of the density plateau the relative density
fluctuations are strongly suppressed, indicating the emergence of
a Mott insulator in this region. (c) Real space density distribution
of the system at different noninteger fillings with $\rho=1/3,\,5/12,\,1/2,\,7/12,\,2/3$,
with the interaction strength kept fixed at $U=15$. One can notice
that density plateaus emerge at all these noninteger fillings. (d)
Corresponding local relative density fluctuation distributions $\widetilde{\Delta}_{j}$
with $\rho=1/3,\,5/12,\,1/2,\,7/12,\,2/3$. See text for more details.}

\label{fig:GS_properties} 
\end{figure}

To investigate the ground state properties of the system we mainly
focus on the local density distributions $\langle\hat{n}_{j}\rangle=\langle\psi_{G}^{R}|\hat{n}_{j}|\psi_{G}^{R}\rangle$
and the local relative density fluctuations $\widetilde{\Delta}_{j}\equiv\sqrt{\langle\hat{n}_{j}^{2}\rangle-\langle\hat{n}_{j}\rangle^{2}}/\langle\hat{n}_{j}\rangle$,
with $|\psi_{G}^{R}\rangle$ being the right ground state of $\hat{H}$.
The density distributions of a half-filled system, i.e., the filling
factor $\rho\equiv N/L=1/2$ with $N$ being the total number of particles,
at different interaction strengths are shown in Fig.~\ref{fig:GS_properties}(a)
(other system parameters are kept fixed with $Je^{\alpha}=1$, $\alpha=2$).
We see that at weak on-site interaction $U/Je^{\alpha}\ll1$, the
nonreciprocal hopping term dominates, and the system shows a strong
NHSE where particles accumulate at the left end of the bosonic chain
{[}cf.~the red curve in Fig.~\ref{fig:GS_properties}(a){]}.

As the on-site repulsive interaction increases, particles on the same
site strongly propel each other hence decreasing the accumulation
of particles at the end of the chain {[}cf.~the blue curve in Fig.~\ref{fig:GS_properties}(a){]}.
At large enough interaction, one expects that the most favorable density
distribution of the system is the one with a region where on average
one particle per site (reducing the on-site energy cost) is located
at one side of the chain (favored by the NHSE). Indeed, as we can
see from the green, black and brown curves in Fig.~\ref{fig:GS_properties}(a),
the system forms a pronounced density plateau once $U>14.2$ (estimated
by the density fluctuation on the first site $\sqrt{\langle\hat{n}_{1}^{2}\rangle-\langle\hat{n}_{1}\rangle^{2}}<0.1$),
where the local relative density fluctuations $\widetilde{\Delta}_{j}$
are strongly suppressed {[}cf. Fig.~\ref{fig:GS_properties}(b){]}.
This thus indicates the emergence of the Mott insulator in this region
despite the filling factor of the system being noninteger {[}$\rho=1/2$
for Figs.~\ref{fig:GS_properties}(a) and \ref{fig:GS_properties}(b){]},
which is in sharp contrast to its Hermitian counterpart, where the
Mott insulator phase can only exist at integer fillings \citep{Sachdev_Book_QPT_2011}. 

From the above discussion, we see that the emergence of a Mott insulator
at $\rho=1/2$ shown in Fig.~\ref{fig:GS_properties}(a) is generally
induced by the competition between NHSE and the on-site repulsive
interaction, therefore we expect Mott insulators could also emerge
at other noninteger fillings. Indeed, as we can see from Figs.~\ref{fig:GS_properties}(c)
and \ref{fig:GS_properties}(d), systems can generally form a Mott
insulator at noninteger fillings ($\rho=1/3,\,5/12,\,1/2,\,7/12,\,2/3$)
in the presence of the strong competition between NHSE and the on-site
repulsive interaction. Moreover, it is worth noting that the density
fluctuations are no longer strongly suppressed in the region away
from the plateau indicating the whole system is not insulating. 

\section{non-Hermitian conservation laws }

Thus far we have investigated the static ground state properties of
the system. Now let us turn to investigate a fundamental aspect of
the dynamical property of the system, namely its conservation laws,
which not only play a crucial role in simplifying and solving physical
problems, but also influence strongly the macroscopic universal dynamical
properties of the system in the vicinities of possible continuous
phase transitions via ``slow modes'' \citep{Hohenberg_RMP_1977}.
For closed quantum systems, this can be usually done by checking the
commutator between the possible candidate and the Hamiltonian operator
of the system under consideration. However, this wisdom no longer
applies to the open quantum systems described by non-Hermitian Hamiltonians.
For instance, the total energy of an open quantum system is generically
not conserved despite the apparent fact that the Hamiltonian operator
of the system commutes with itself. This thus leads us first to investigate
the general requirement for any observable being a conserved quantity
in generic non-Hermitian quantum systems.

\subsection{General requirement of non-Hermitian conservation laws }

We first denote the Hermitian operator associated with a generic physical
observable by $\hat{O}$, whose expectation value at time $t$ is
denoted as $\bar{O}(t)\equiv\langle\psi(t)|\hat{O}|\psi(t)\rangle$,
with $|\psi(t)\rangle$ being the state of a generic time-independent
non-Hermitian system described by the Hamiltonian $\hat{H}$. The
formal solutions of $|\psi(t)\rangle$ and $\langle\psi(t)|$ are
$|\psi(t)\rangle=e^{-i\hat{H}t}|\psi(t=0)\rangle$ (units are chosen
such that $\hbar=1$) and $\langle\psi(t)|=\langle\psi(t=0)|e^{i\hat{H}^{\dagger}t}$,
respectively. If $\hat{O}$ is a conserved quantity, then $\partial_{t}\bar{O}(t)=i\langle\psi(0)|e^{i\hat{H}^{\dagger}t}(\hat{H}^{\dagger}\hat{O}-\hat{O}\hat{H})e^{-i\hat{H}t}|\psi(0)\rangle=0$
have to be satisfied, indicating $\hat{O}$ have to satisfy the requirement
\begin{equation}
\hat{H}^{\dagger}\hat{O}-\hat{O}\hat{H}=0,\label{eq:conservation_condition_non-Hermitian}
\end{equation}
where we immediately notice that any Hermitian operator $\hat{O}$
that commutes with the Hamiltonian operator, i.e., $[\hat{O},\hat{H}]=0$,
is guaranteed to be \emph{not conserved} due to the non-Hermiticity,
i.e., $\hat{H}^{\dagger}\hat{O}-\hat{O}\hat{H}=(\hat{H}^{\dagger}-\hat{H})\hat{O}\neq0$.
This is in sharp contrast to the Hermitian quantum systems, where
any Hermitian operator that commutes with the Hamiltonian operator
is guaranteed to be a conserved quantity, and raises the fundamental
question of the existence of conserved laws in non-Hermitian quantum
systems and their concrete forms. Indeed, as we shall present in the
following, such conserved quantities indeed exist and can be systematically
constructed for non-Hermitian systems with completely real energy
spectra.

We first notice that for any non-Hermitian diagonalizable Hamiltonian
$\hat{H}$ with a completely real spectrum, one can always construct
a similarity transformation $\hat{S}$ which transforms $\hat{H}$
into an auxiliary \emph{Hermitian} Hamiltonian $\hat{H}_{\mathrm{aux}}$
with the same spectrum (cf.~Ref.~\citep{Mostafazadeh_J_Math_Phys_2002}
or a short derivation presented in the Appendix), i.e., $\hat{S}^{-1}\hat{H}\hat{S}=\hat{H}_{\mathrm{aux}}$.
Therefore, one could reformulate Eq.~(\ref{eq:conservation_condition_non-Hermitian})
in terms of the similarity transformation $\hat{S}$ and the corresponding
auxiliary \emph{Hermitian} Hamiltonian, i.e., $(\hat{S}^{\dagger})^{-1}\hat{H}_{\mathrm{aux}}\hat{S}^{\dagger}\hat{O}-\hat{O}\hat{S}\hat{H}_{\mathrm{aux}}\hat{S}^{-1}=0$,
which is equivalent to 
\begin{equation}
[\hat{S}^{\dagger}\hat{O}\hat{S},\hat{H}_{\mathrm{aux}}]=0,
\end{equation}
indicating if $\hat{S}^{\dagger}\hat{O}\hat{S}$ is a conserved quantity,
denoted as $\hat{C}_{\mathrm{aux}}$, of the auxiliary Hermitian Hamiltonian
$\hat{H}_{\mathrm{aux}}$, then $\hat{O}=(\hat{S}^{\dagger})^{-1}\hat{C}_{\mathrm{aux}}\hat{S}^{-1}$
is a conserved quantity of the non-Hermitian system, i.e., 
\begin{align}
 & \partial_{t}\bar{O}(t)=0\,\mathrm{for}\,\hat{O}=(\hat{S}^{\dagger})^{-1}\hat{C}_{\mathrm{aux}}\hat{S}^{-1},\,\mathrm{if}\,[\hat{C}_{\mathrm{aux}},\hat{H}_{\mathrm{aux}}]=0.\label{eq:non-Hermitian_conservation_laws}
\end{align}
This thus provides a systematic way to identify the conservation laws
in non-Hermitian quantum systems. In the following, we shall use the
interacting bosonic chain with nonreciprocal hopping described by
the Hamiltonian (\ref{eq:Hamiltonian}) as a concrete example to illustrate
identifying the conservation laws in non-Hermitian quantum systems.

\subsection{Non-Hermitian conservation laws in an interacting bosonic chain with
nonreciprocal hopping}

\begin{figure}
\includegraphics[width=1.65in]{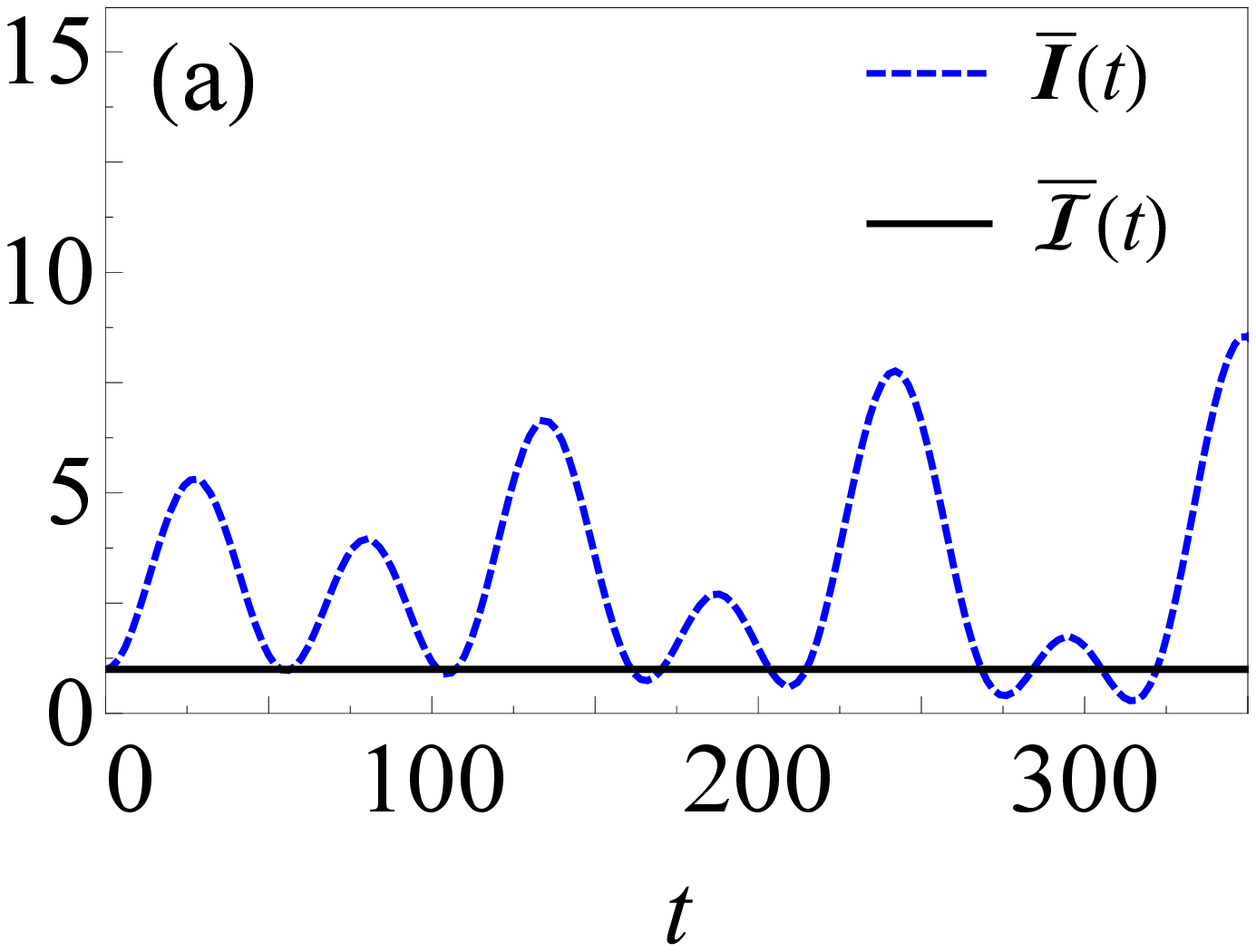}~~\includegraphics[width=1.6in]{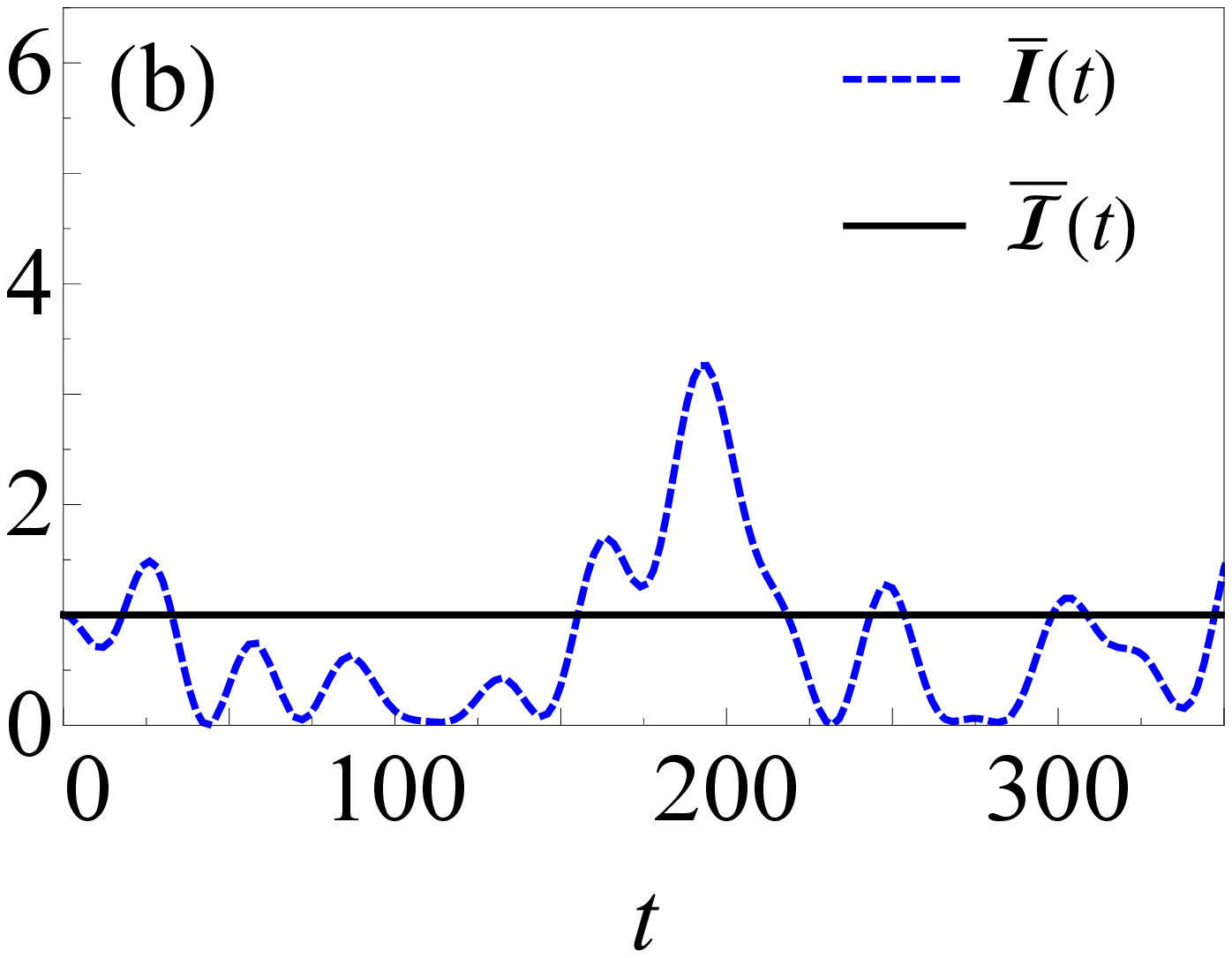}

\includegraphics[width=1.65in]{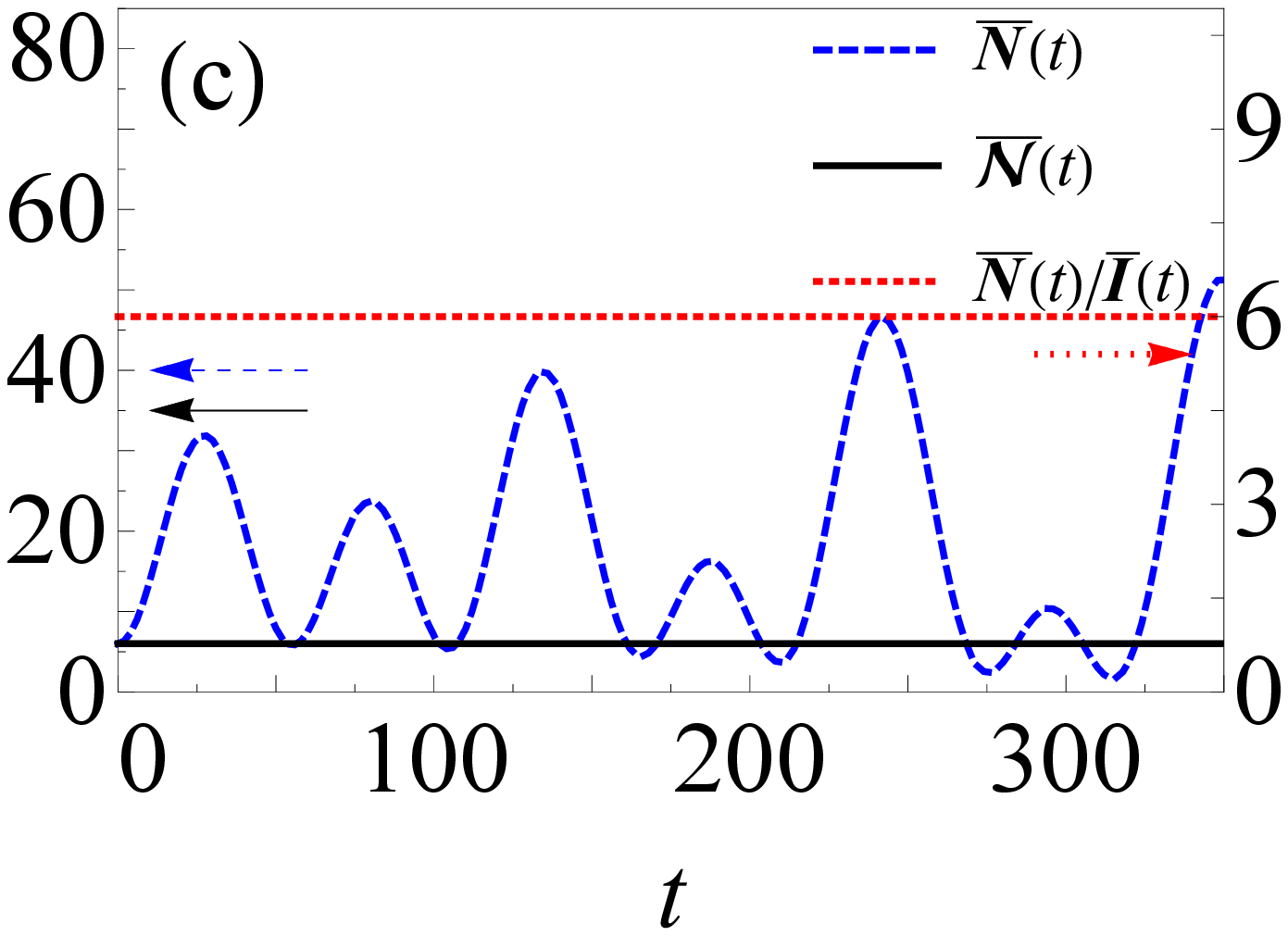}~\includegraphics[width=1.65in]{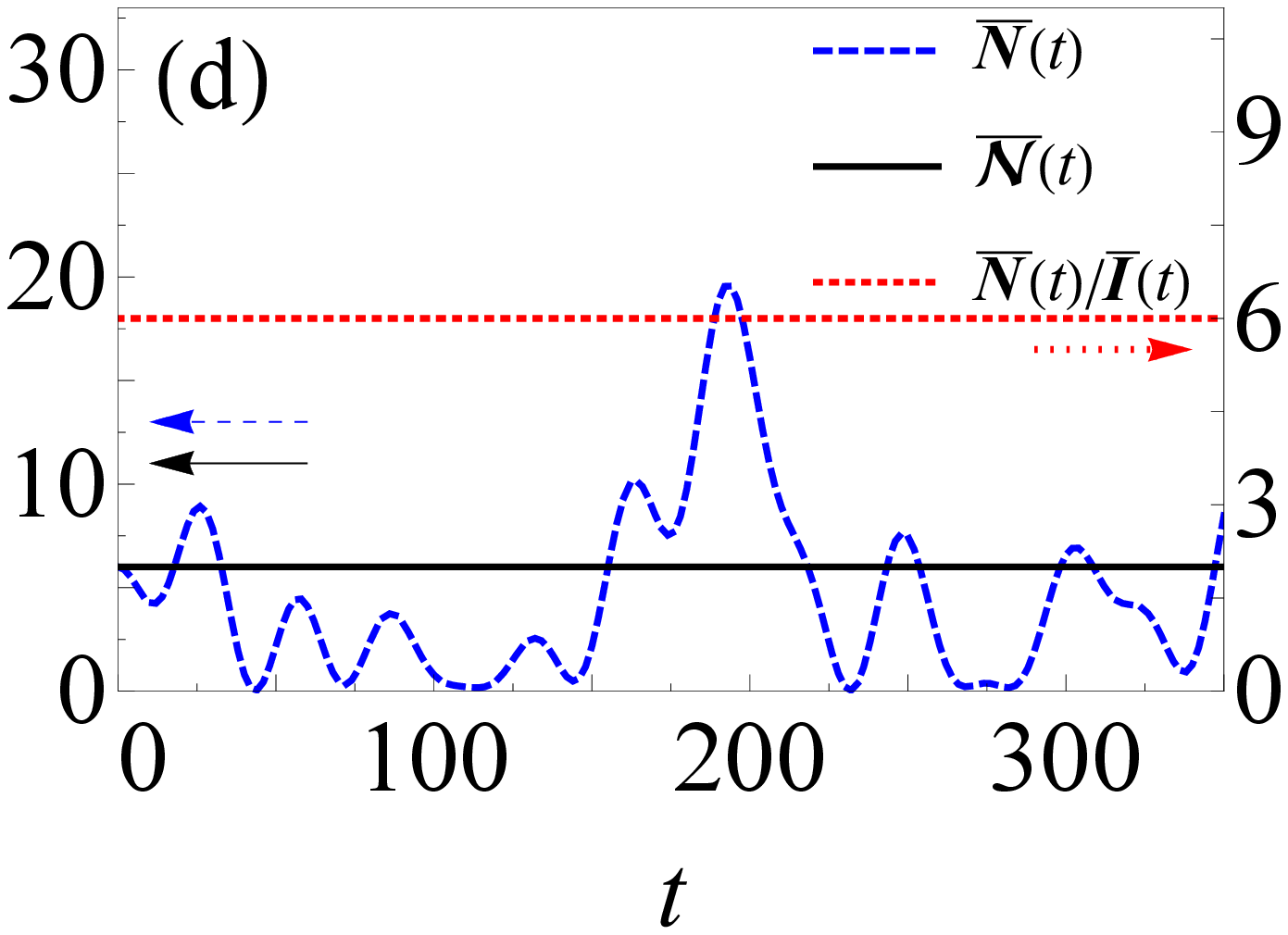} 

\includegraphics[width=1.65in]{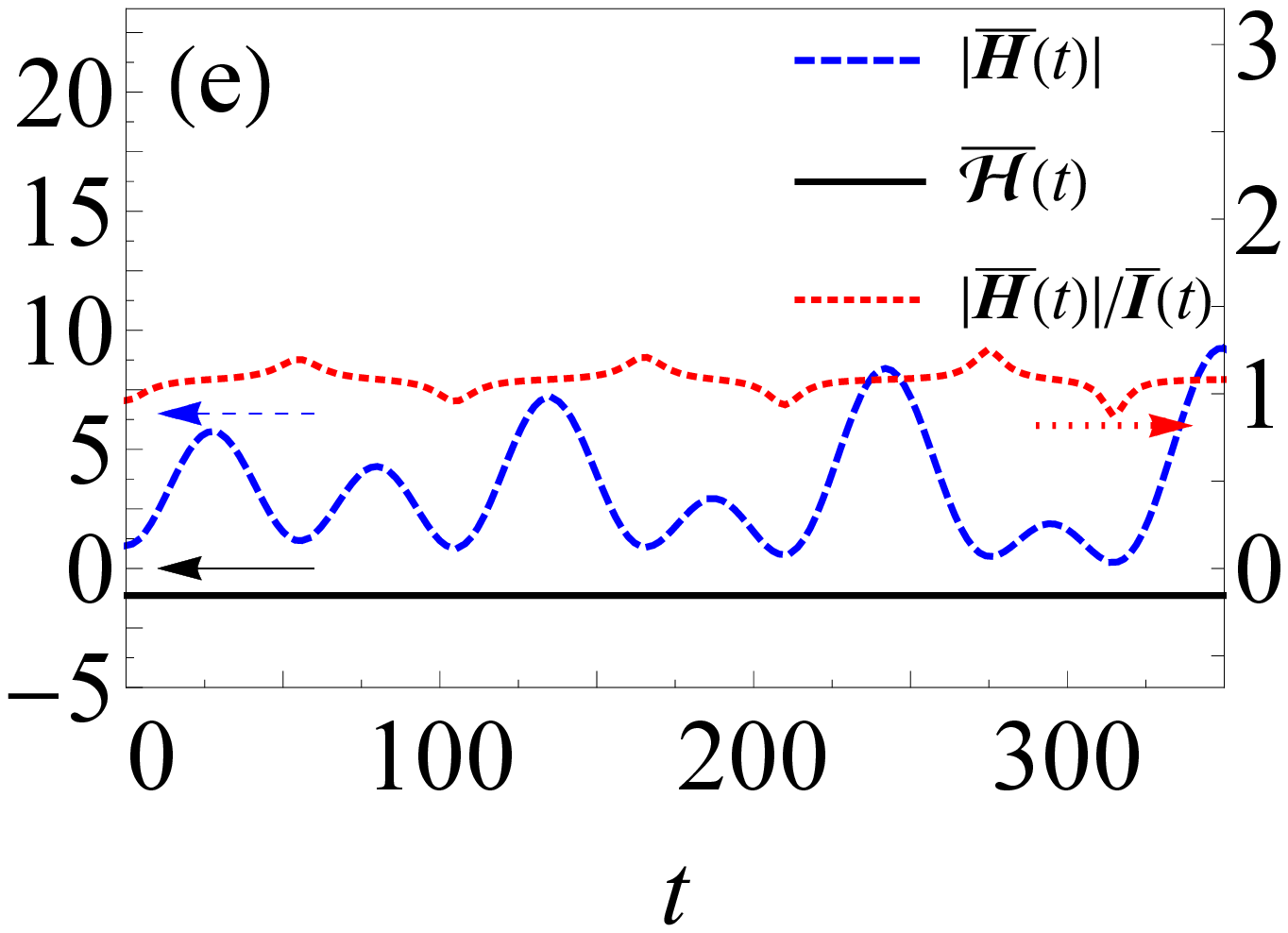}~\includegraphics[width=1.65in]{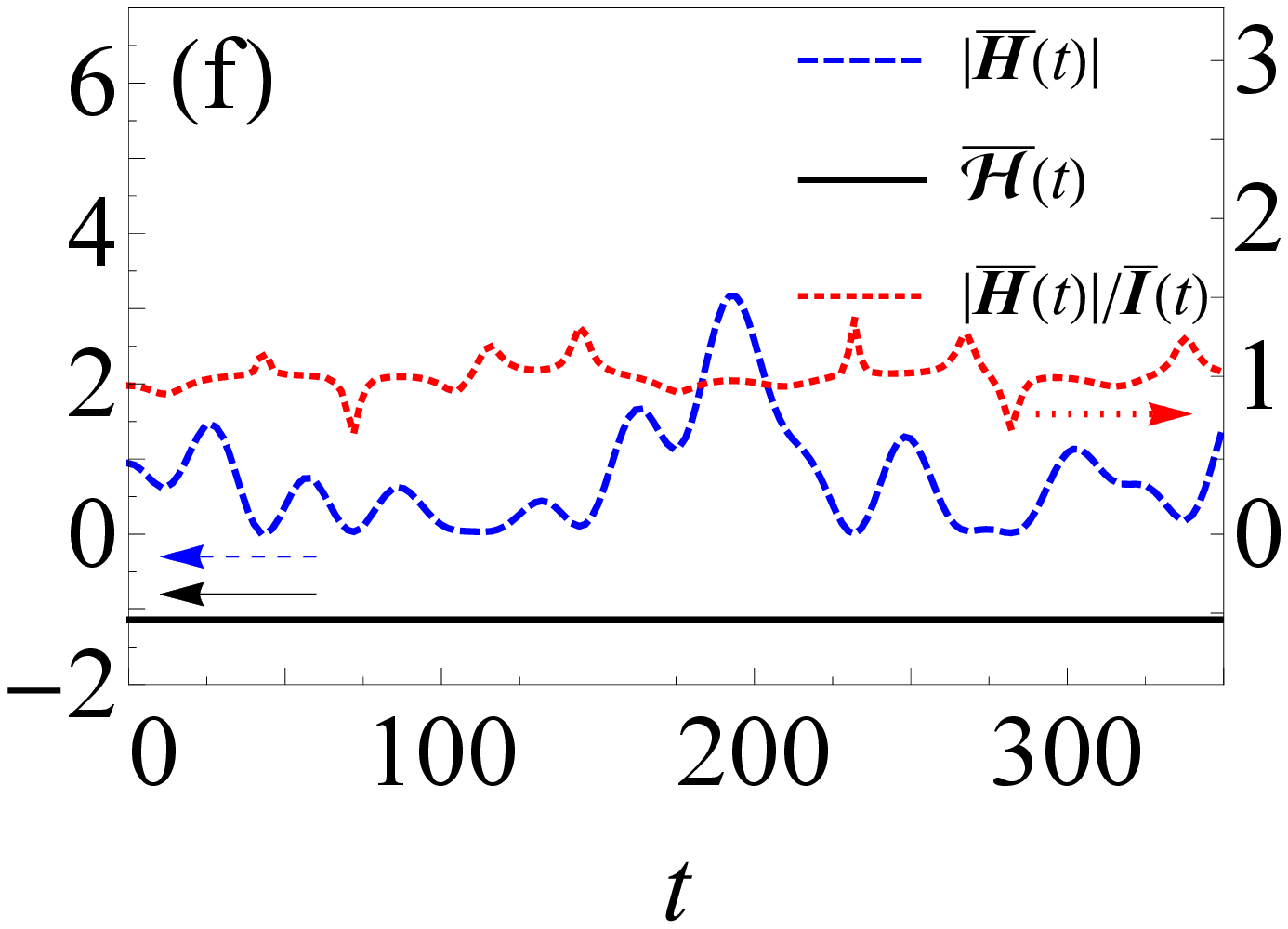}

\caption{Quantum dynamics of the expectation values for the conserved observables
$\hat{\mathcal{I}}$, $\hat{\mathcal{N}}$, $\hat{\mathcal{H}}$ and
the related operators $\hat{I}$, $\hat{N}$, $\hat{H}$ that commute
with the system's Hamiltonian. The parameters of the system are $\alpha=2$,
$Je^{\alpha}=1$, $U=1$, $L=12$. The \textquotedblleft gauge\textquotedblright{}
constant $A$ is chosen in such a way that makes $\bar{\mathcal{I}}(t=0)=1$.
(a), (c), (e) Quantum dynamics of the system with the initial state
$|\psi(t=0)\rangle=\left|\sum_{m,n=1}^{M}\langle\psi_{n}^{R}|\psi_{m}^{R}\rangle\right|^{-1/2}\sum_{m=1}^{M}|\psi_{m}^{R}\rangle$
with $M=3$ and the total particle number being 6. (b), (d), (f) Quantum
dynamics of the system with the initial state being the same as the
one for (a), (c), (e) except $M=10$. See text for more details. }
\label{fig:Dynamics} 
\end{figure}

As one can notice from Eq.~(\ref{eq:non-Hermitian_conservation_laws}),
as long as one finishes the construction of the similarity transformation
$\hat{S}$, identification of the non-Hermitian conservation laws
directly follows. For the bosonic non-Hermitian chain with nonreciprocal
hopping under consideration, one can take the similarity transformation
employed in the investigation of the ``non-Hermitian skin effect''
\citep{Yao_PRL_2018} as a convenient choice for $\hat{S}$, whose
explicit form in the second quantization form reads $\hat{S}=A\exp(-\alpha\sum_{j=1}^{L}j\hat{n}_{j})$,
with $A$ being a nonzero ``gauge'' constant. One can straightforwardly
show that the similarity transformation $\hat{S}$ can transform the
non-Hermitian Hamiltonian (\ref{eq:non-Hermitian_conservation_laws})
into the Hermitian Hamiltonian of the conventional Bose-Hubbard chain,
i.e., $\hat{S}^{-1}\hat{H}\hat{S}=\hat{H}_{\mathrm{BH}}$ with $\hat{H}_{\mathrm{BH}}=-J\sum_{j=1}^{L-1}(\hat{b}_{j}^{\dagger}\hat{b}_{j+1}+\mathrm{H.c.})+\frac{U}{2}\sum_{j=1}^{L}\hat{n}_{j}(\hat{n}_{j}-1)$.
From the form of $\hat{H}_{\mathrm{BH}}$, one can directly notice
that the total particle number operator $\hat{N}\equiv\sum_{j=1}^{L}\hat{n}_{j}$,
$\hat{H}_{\mathrm{BH}}$ itself, and the identity operator $\hat{I}$
commute with $\hat{H}_{\mathrm{BH}}$, corresponding to the total
particle number, energy, probability conservation in the conventional
Bose-Hubbard chain, respectively. Via Eq.~(\ref{eq:non-Hermitian_conservation_laws}),
we thus can directly identify three conservation laws for the non-Hermitian
system, namely, the observables that correspond to the Hermitian operators,
\begin{align}
\hat{\mathcal{N}} & \equiv(\hat{S}^{\dagger})^{-1}\hat{N}\hat{S}^{-1},\label{eq:CQ_deformed_N}\\
\hat{\mathcal{H}} & \equiv(\hat{S}^{\dagger})^{-1}\hat{H}_{\mathrm{BH}}\hat{S}^{-1},\label{eq:CQ_deformed_H_BH}\\
\hat{\mathcal{I}} & \equiv(\hat{S}^{\dagger})^{-1}\hat{I}\hat{S}^{-1},\label{eq:CQ_deformed_I}
\end{align}
are conserved.

To further corroborate the above analytic results, we numerically
simulate the quantum dynamics of the system and calculate the time
evolution of the expectation value of the conserved quantities, i.e.,
$\bar{\mathcal{N}}(t)$, $\bar{\mathcal{H}}(t)$, $\bar{\mathcal{I}}(t)$,
and also those of the related operators that \emph{commute} with the
Hamiltonian $\hat{H}$ of the system, i.e., $\bar{N}(t)$, $\bar{H}(t)$,
$\bar{I}(t)$. More specifically, as shown in Fig.~\ref{fig:Dynamics},
we choose two initial states of the form $|\psi(t=0)\rangle=\left|\sum_{m,n=1}^{M}\langle\psi_{n}^{R}|\psi_{m}^{R}\rangle\right|^{-1/2}\sum_{m=1}^{M}|\psi_{m}^{R}\rangle$
with $M=3,10$ and the total particle number being 6 ($|\psi_{m}^{R}\rangle$
is the $m$th lowest right eigenstate of $\hat{H}$) and simulate
the corresponding quantum dynamics of the system with $L=12$, $U=1$,
$\alpha=2$, $Je^{\alpha}=1$ kept fixed. As one can see from Fig.~\ref{fig:Dynamics},
while $\bar{N}(t)$, $\bar{H}(t)$, $\bar{I}(t)$ manifest a strong
time dependence despite the fact that their corresponding operators
\emph{commute} with the Hamiltonian {[}it is interesting to notice
that $\bar{N}(t)/\bar{I}(t)$ manifests a time-independent behavior,
however this mainly originates from the fact that the total particle
number operator $\hat{N}$ commutes with the similarity transformation
operator{]}, $\bar{\mathcal{N}}(t)$, $\bar{\mathcal{H}}(t)$, $\bar{\mathcal{I}}(t)$
indeed remain exactly at their constant values during the whole quantum
dynamics. 

Finally, we would like to remark on a crucial difference between the
conservation laws in non-Hermitian and Hermitian systems. In non-Hermitian
systems, the existence of conservation laws manifests a strong sensitivity
on the system's boundary condition, since its change can turn the
completely real spectrum into a complex one in certain cases, which
consequentially excludes the existence of the conserved quantity.
For instance, systems described by noninteracting or interacting Hamiltonians
similar to Hamiltonian (\ref{eq:Hamiltonian}) with periodic boundary
conditions imposed assume complex energy spectra \citep{Yao_PRL_2018,Zhang_PRB_2020}

\section{Experimental observability}

\begin{figure}
\includegraphics[width=1.65in]{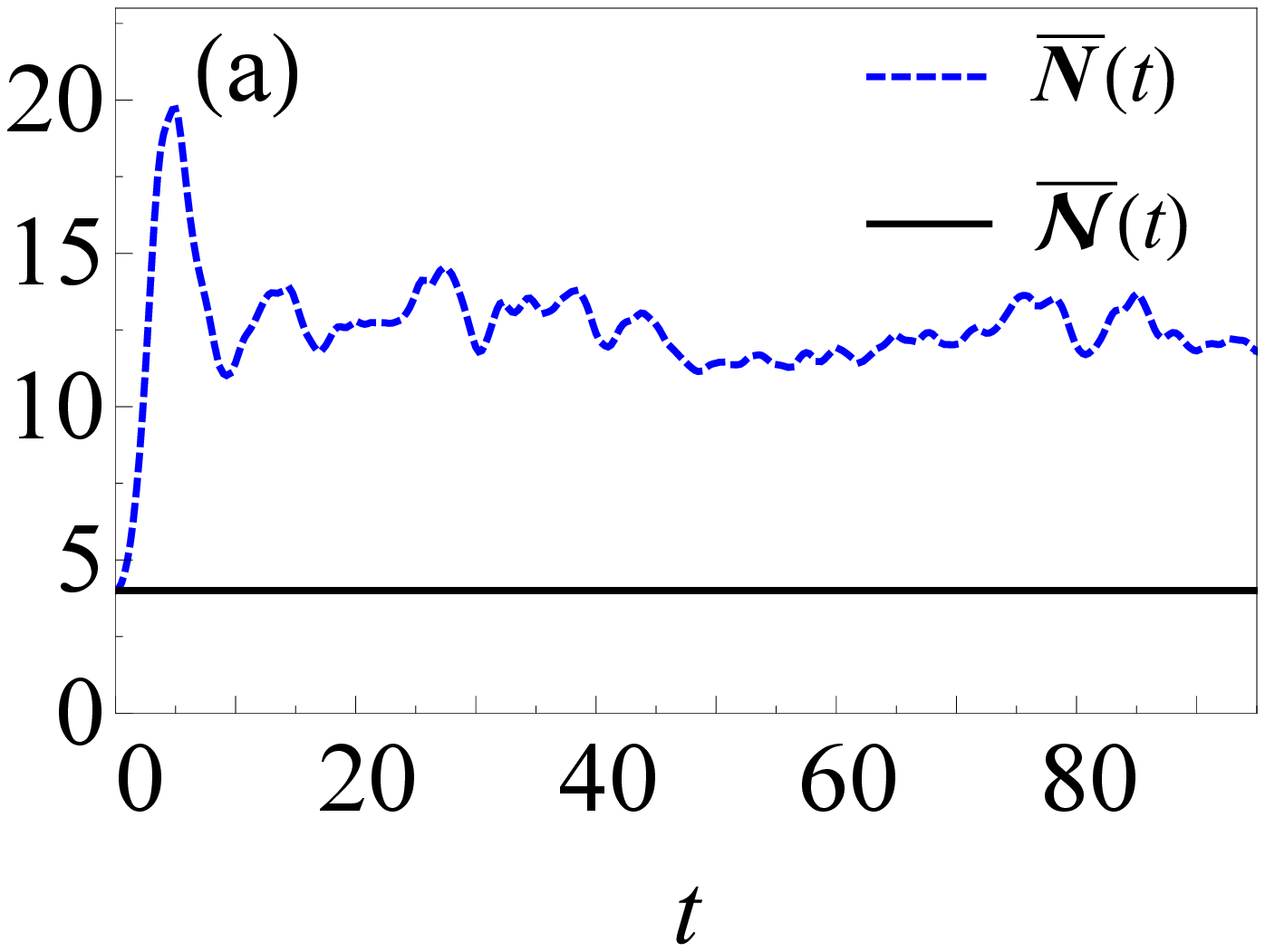}~\includegraphics[width=1.65in]{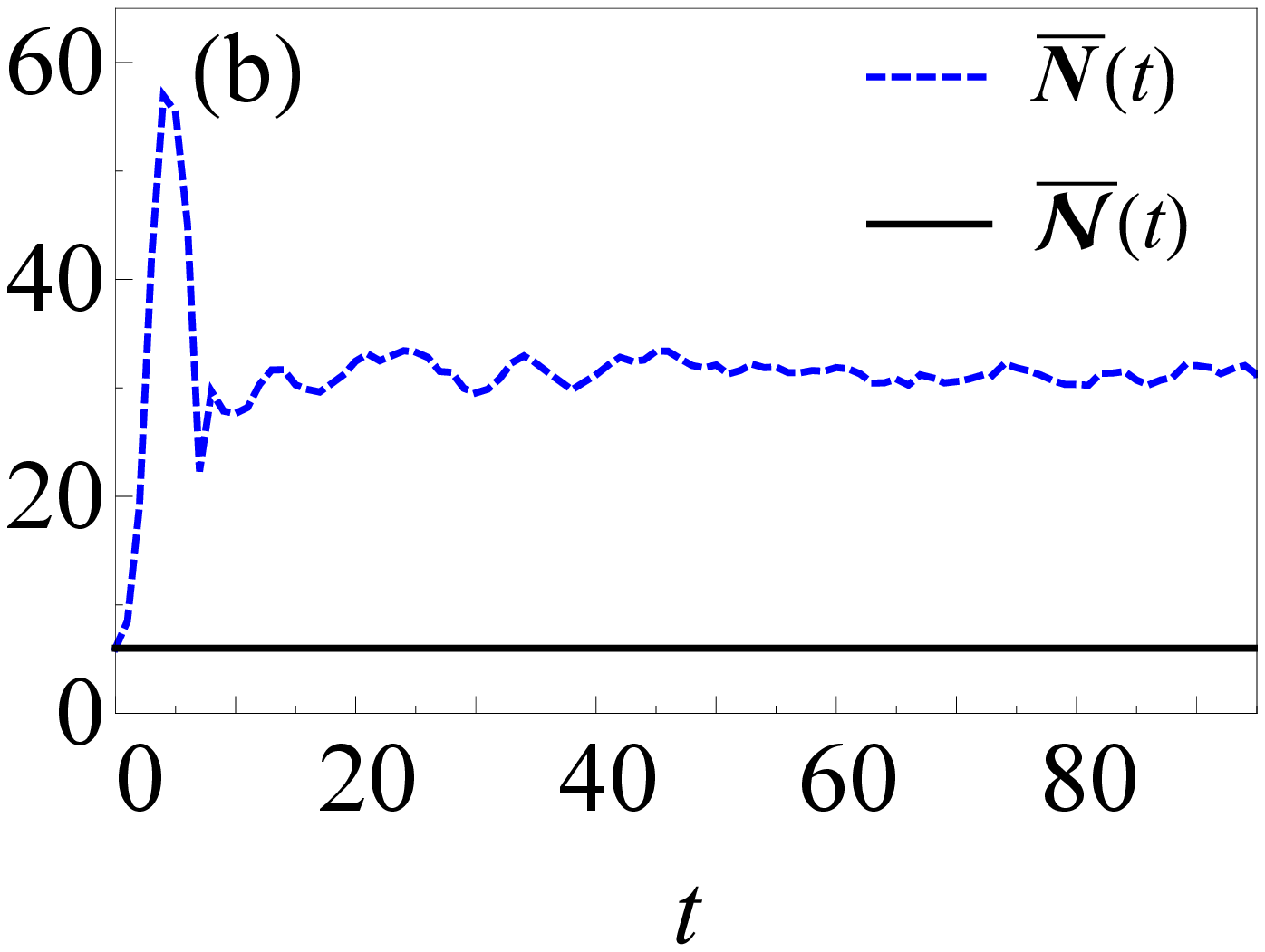}\caption{Quantum dynamics of the expectation value of the conserved quantity
$\mathcal{\hat{N}}$ and also the total particle number $\hat{N}$
with centrally filled Fock states as the initial states. System parameters
are $\alpha=0.1$, $Je^{\alpha}=1$, $U=1$, $L=12$. The \textquotedblleft gauge\textquotedblright{}
constant $A$ is chosen in such a way that makes $\bar{\mathcal{I}}(t=0)=1$.
(a) Time evolutions of $\bar{\mathcal{N}}(t)$ and $\bar{N}(t)$ with
$|\psi(t=0)\rangle=|000011110000\rangle$ being the initial state.
(b) Time evolutions of $\bar{\mathcal{N}}(t)$ and $\bar{N}(t)$ with
$|\psi(t=0)\rangle=|000111111000\rangle$ being the initial state.
See text for more details.}
\label{fig:Exp_observability} 
\end{figure}

We expect that the Mott insulators in systems with noninteger filling
and the non-Hermitian conservation laws identified in this work can
be observed experimentally. For instance, by employing the well-established
experimental platform offered by ultracold atoms in optical lattices
\citep{Greiner_Nature_2002,Bloch_RMP_2008}, the nonreciprocal hopping
can be effectively engineered by introducing atom loss \citep{Yan2020,Li_PRL_2020,RenJo2021,Li_Nat_Com_2019}.
The direct observation of the Mott insulator in systems with noninteger
filling can be readily performed via measuring the local density distribution
with a quantum gas microscope \citep{Bakr_Nature_2009}. For the non-Hermitian
conservation laws, we expect the one associated with $\hat{\mathcal{N}}$
is particularly feasible to be observed in current experimental setups
via measuring the local density $\hat{n}_{j}$ with a quantum gas
microscope at different time points during the quantum dynamics, and
reconstructing the expectation value of $\hat{\mathcal{N}}$ according
to Eq.~(\ref{eq:CQ_deformed_N}). For instance, one could choose
the Fock state with the central part of the system uniformly filled
as the initial state which is most accessible in experiments, and
monitor the time evolution of $\bar{\mathcal{N}}(t)$. In Fig.~\ref{fig:Exp_observability},
the time evolutions of $\bar{\mathcal{N}}(t)$ {[}also $\bar{N}(t)${]}
with two centrally filled Fock states ($|000011110000\rangle$ and
$|000111111000\rangle$) of the system with $L=12$ as the initial
states are shown. As we can see, while the direct measurement of the
total particle number $\bar{N}(t)$ is expected to manifest a strong
time dependence, the experimental measurements of $\bar{\mathcal{N}}(t)$
are expected to remain unchanged at different time points.

\section{Conclusions}

Non-Hermiticity together with interactions can give rise to rich many-body
physics beyond those of corresponding Hermitian systems, as interacting
bosonic chains with nonreciprocal hopping have shown: The systems
with noninteger filling can support Mott insulators due to the collaboration
between nonreciprocal hopping and the on-site interaction. The conservation
laws for non-Hermitian systems generically manifest a stark difference
from their Hermitian counterpart, especially for any Hermitian operator
that commutes with the Hamiltonian operator, its expectation value
is guaranteed to be \emph{nonconserved} in the non-Hermitian quantum
dynamics. The non-Hermitian conservation laws can be systematically
constructed in non-Hermitian many-body quantum systems with a completely
real spectrum via the auxiliary Hermitian system. We believe our work
will stimulate further theoretical and experimental investigations
on interacting non-Hermitian many-body systems, in particular their
quantum dynamical properties. 
\begin{acknowledgments}
We thank Xiancong Lu, Guoqing Zhang, and Danwei Zhang for helpful
discussions. This work was supported by NSFC (Grants No.~11874017
and No.~11904109), GDSTC under Grant No.~2018A030313853, Guangdong
Basic and Applied Basic Research Foundation (Grant No.~2019A1515111101),
Science and Technology Program of Guangzhou (Grant No.~2019050001),
and START grant of South China Normal University. 
\end{acknowledgments}

\appendix
%dummy comment inserted by tex2lyx to ensure that this paragraph is not empty

\section{Similarity transformations between non-Hermitian Hamiltonians with
completely real spectra and Hermitian Hamiltonians \label{sec:Math_properties_Non-Hermitian-Hamiltonian}}

To be self-contained, we review here a few mathematical properties
of non-Hermitian diagonalizable Hamiltonians with completely real
spectra presented in Ref.~\citep{Mostafazadeh_J_Math_Phys_2002}.

For a non-Hermitian Hamiltonian $\hat{H}$ with a completely real
spectrum $E_{n}$, it assumes the formal form 
\begin{equation}
\hat{H}=\sum_{n}E_{n}|\psi_{n}^{R}\rangle\langle\psi_{n}^{L}|,\label{eq:non-Hermitian_H_formal_form}
\end{equation}
where $|\psi_{n}^{R}\rangle$ and $|\psi_{n}^{L}\rangle$ are the
right and left eigenstates of $\hat{H}$, respectively. These two
sets of eigenstates satisfy the following relations,
\begin{align}
 & \hat{H}|\psi_{n}^{R}\rangle=E_{n}|\psi_{n}^{R}\rangle,\,\hat{H}^{\dagger}|\psi_{n}^{L}\rangle=E_{n}|\psi_{n}^{L}\rangle,\\
 & \langle\psi_{m}^{L}|\psi_{n}^{R}\rangle=\delta_{mn},\,\sum_{n}|\psi_{n}^{R}\rangle\langle\psi_{n}^{L}|=\hat{I},
\end{align}
with $\delta_{mn}$ being the Kronecker delta function and $\hat{I}$
being the identity operator of the Hilbert space $\mathbb{H}$ associated
with $\hat{H}$. Since the spectrum of the non-Hermitian Hamiltonian
$\hat{H}$ is real, one can always construct a auxiliary Hermitian
Hamiltonian $\hat{H}_{\mathrm{aux}}$ that shares the same spectrum
with $\hat{H}$, i.e., 
\begin{equation}
\hat{H}_{\mathrm{aux}}\equiv\sum_{n}E_{n}|n\rangle\langle n|,\label{eq:hermitian_H_aux}
\end{equation}
where $\{|n\rangle\}$ is a complete set of the orthonormal basis
of the Hilbert space $\mathbb{H}$, i.e., $\sum_{n}|n\rangle\langle n|=\hat{I},$
$\langle m|n\rangle=\delta_{mn}$. By comparing Eqs.~(\ref{eq:non-Hermitian_H_formal_form})
and (\ref{eq:hermitian_H_aux}), one can notice that $\hat{H}$ can
be transformed to $\hat{H}_{\mathrm{aux}}$ via a similarity transformation,
i.e., 
\begin{equation}
\hat{S}^{-1}\hat{H}\hat{S}=\hat{H}_{\mathrm{aux}},
\end{equation}
with 
\begin{equation}
\hat{S}\equiv\sum_{n}|\psi_{n}^{R}\rangle\langle n|,\,\hat{S}^{-1}=\sum_{n}|n\rangle\langle\psi_{n}^{L}|.
\end{equation}

\end{document}